%% file: combined.tex
\documentclass{llncs}
\usepackage{amssymb,latexsym,epsfig,algorithmic,algorithm,wrapfig}
\usepackage{subfigure,fullpage}

\begin{document}

\mainmatter

\title{Pancake Flipping with Two Spatulas}
%\title{Sorting by Prefix Reversals and Prefix Transpositions}
%\titlerunning{Sorting by Prefix Reversals and Prefix Transpositions}

%\author{Mahfuza Sharmin\inst{1,3} \and Rukhsana Yeasmin\inst{1,3} \and Masud Hasan\inst{1,4} \and Atif Rahman\inst{1,5} \and M. Sohel Rahman\inst{1,2,4}}
\author{Masud Hasan\inst{1,3} \and Atif Rahman\inst{1,4} \and M. Sohel Rahman\inst{1,2,3} \and Mahfuza Sharmin\inst{1,5} \and Rukhsana Yeasmin\inst{1,5}}

%\institute{Department of Computer Science and Engineering\\
%Bangladesh University of Engineering and Technology, Dhaka-1000, Bangladesh\\
%Email: \email\{mhfz\_sharmin, smrity\_23\}@yahoo.com,
%masudhasan@cse.buet.ac.bd, atif.bd@gmail.com, sohel\_rahman\_joy@yahoo.com}

\institute{Department of Computer Science and Engineering\\Bangladesh University of Engineering and Technology, Dhaka-1000, Bangladesh\\
%\texttt{http://www.buet.ac.bd/cse}
\and
Algorithm Design Group, Department of Computer Science\\King's College London, Strand, London WC2R 2LS, England\\
%\texttt{http://www.dcs.kcl.ac.uk/adg}\\
\and \email{\{masudhasan, msrahman\}@cse.buet.ac.bd}
\and \email{atif.bd@gmail.com}
\and \email{\{mhfz\_sharmin, smrity\_23\}@yahoo.com}}

%\authorrunning{Hasan, Rahman, Rahman, Sharmin and Yeasmin}

\maketitle

%----------------------------------------------------------------------------
\begin{abstract}
In this paper we study several variations of the \emph{pancake flipping problem}, 
which is also well known as the problem of \emph{sorting by prefix reversals}.
We consider the variations in the sorting process by adding with prefix reversals other similar operations such as
prefix transpositions and prefix transreversals.
These type of sorting problems have applications in interconnection networks and computational biology.
We first study the problem of sorting unsigned permutations by prefix reversals and prefix transpositions
and present a 3-approximation algorithm for this problem.
Then we give a 2-approximation algorithm for sorting by prefix reversals and prefix transreversals. 
We also provide a 3-approximation algorithm for sorting by prefix reversals and prefix 
transpositions where the operations are always applied at the unsorted suffix of the permutation.
We further analyze the problem in more practical way and show quantitatively how 
approximation ratios of our algorithms improve with the increase of number of prefix
reversals applied by optimal algorithms. 
Finally, we present experimental results to support our analysis.

\end{abstract}
%------------------------------------------------------------------------
\section{Introduction}
\label{se:intro}

Given a permutation $\pi$, a \emph{reversal} reverses a substring of $\pi$,
a \emph{transposition} cuts a substring of $\pi$ and pastes it in a different location,
and a \emph{transreversal} is a transposition of a substring with a reversal done before it is pasted.
In a \emph{prefix} reversal/transposition/transreversals the corresponding substring
is always a prefix of $\pi$.

The \emph{pancake flipping problem}~\cite{amm,BS03,HP95,HS93,HS97} deals with 
finding the minimum number of prefix reversals (i.e., \emph{flips}) required to
sort a given permutation.
This problem was first introduced in 1975 by~\cite{amm} which describes
the motivation of a chef to rearrange a stack of pancakes from the smallest 
pancake on the top to the largest one on the bottom by grabbing several pancakes 
from the top with his spatula and flipping them over, repeating them as many times 
as necessary.

Aside from being an interesting combinatorial problem, this problem and its variations
have applications in interconnection networks and computational biology.
The number of flips required to sort the stack of $n$ pancakes is the \emph{diameter} 
of the $n$-dimensional \emph{pancake network}~\cite{HS93,HS97}.
The diameter of a network is the maximum distance between any pair of nodes in the network
and corresponds to the worst communication delay for broadcasting messages in the network~\cite{HS93,HS97}.
A well studied variation of pancake flipping problem is the \emph{burnt pancake flipping problem}~\cite{BS03,HS93,HS97}
where each element in the permutation has a sign, and the sign of an element changes with reversals. 
Pancake and burnt pancake networks have better diameter and better vertex degree than the popular 
hypercubes~\cite{HS93}.
There exists some other variations of pancake flipping, giving different efficient interconnection networks~\cite{BS03}. 

A broader class consisting of similar sorting problems, called the \emph{genome rearrangement problems},
are extensively studied in computational molecular biology,
%Genome rearrangement problems are studied to analyze differences in gene orders in genomes of two or more species.
%In genome rearrangement problems 
where the orders of genes in two species
are represented by permutations and the problem is to transform one into another
by using minimum number of pre-specified rearrangement operations.
%The problem is polynomially equivalent to sorting a permutation using
%the set of pre-specified rearrangement operations.
%
In order to explain the existence of essentially the same set of genes but differences 
in their order in different species, several rearrangement operations have been suggested,
including reversals~\cite{KS93,BP93,BHK02}, block interchange~\cite{BI96},
transpositions~\cite{BP95,EH05,H03}, transreversals~\cite{HS04}, fission and fusion~\cite{FF06}, 
prefix transposition~\cite{DM02}, etc. 

The abovementioned sorting problems are mostly NP-complete or their complexity is unknown.
Caprara~\cite{C97} proved that sorting by reversals is NP-hrad,
whereas Heydari and Sudborough~\cite{HS} have claimed that 
sorting by prefix reversals is NP-complete too.
The complexity of sorting by transposition and sorting by prefix transposition is still open.
As a result, many approximation algorithms are known for each of these problems and their variations.

A number of authors have also considered the problem of sorting permutations by using 
more than one rearrangement operations (reversals, transpositions etc.)~\cite{W98,GPS99,LX99,HS04,E02,Atif08,LZ08},
mostly for signed permutations.
Rahman et.al.~\cite{Atif08} studied the problem of sorting permutations by transpositions and reversals,
where they give an approximation algorithm with approximation ratio $2.83$.

\subsection{Our results}
In this paper we study some variations of pancake flipping problem
from the view point of sorting permutations.
We consider the problem of sorting an (unsigned) permutation by prefix reversals and prefix transpositions. 
We give a 3-approximation algorithm for sorting by prefix reversals 
and prefix transpositions, and a 2-approximation algorithm for sorting by prefix reversals 
and prefix transreversals.
%, where a (prefix) transreversal is a (prefix) transposition 
%with the subsequence reversed before it is pasted into a new location. 
Experimental result shows that our algorithms perform much better in practice.

Note that the problem of sorting by reversals and transpositions in~\cite{Atif08}
and the problem of sorting by prefix reversals and prefix transpositions considered
in this paper are not the same and they do not imply each other.

We also introduce the concept of \emph{forward march}. The idea of forward 
march comes naturally from a greedy approach where someone may try to sort the permutation
from starting to end. While applying prefix
reversals and prefix transpositions, a prefix of the given
permutation may become sorted. Whenever this happens, we move
forward and apply the next operation in the remaining unsorted
suffix of the permutation. We give a 3-approximation algorithm for this problem 
which also performs better on average.

%The waiter described in~\cite{amm} may find an use for the above problems that we consider in this paper
%to rearrange the pancakes in order by size if he has two free hands. 
The above problems that we consider in this paper are variations of the original
pancake flipping problem where the chef has two spatulas in his two free hands.
%may come to a use to the chef described in~\cite{amm}
%who wants to rearrange the pancakes in order by size if he has two free hands. 
He can either lift some pancakes from the top of the stack and flip them (a prefix reversal)
or he can lift a top portion of the stack with one hand, lift another portion from the top with the other hand, 
and place 
%them in opposite order 
the top portion under the second portion (a prefix transposition)
possibly with a flip (a prefix transreversal).
Also, time to time, when a top portion of the stack is sorted he can remove it from the stack (a forward march).

It is worth mentioning that the worst case ratios of our algorithms can only be realized
when an optimal algorithm applies no prefix reversals at all. But it is very likely that
an optimal algorithm will apply both operations.
Keeping this observation in mind, we derive mathematically the equations 
for approximation ratio in terms of the number of prefix reversals applied by 
an optimal algorithm. 

%\subsection{Outline}
We organize rest of the paper as follows. In Section~\ref{se:pre} we give
the definitions and other preliminaries. In Section~\ref{se:tr3},~\ref{se:tr2}, and~\ref{se:march} 
we present the approximation algorithms. 
In Section~\ref{se:alt} we derive equations for approximation 
ratio in terms of number of prefix reversals applied by an optimal algorithm.
In Section~\ref{se:exp} we present our experimental results.  
Finally, Section~\ref{se:con} concludes the paper.
%----------------------------------------------------------------------
%**********************************************************************
\section{Preliminaries}
\label{se:pre}

Let $\pi=[\pi_0,\pi_1,\ldots,\pi_n,\pi_{n+1}]$ be a permutation of
$n+2$ distinct elements where $\pi_0=0$, $\pi_{n+1}=n+1$ and $1\leq
\pi_i\leq n$ for each $1\leq i\leq n$ (the middle $n$ elements of
$\pi$ are to be sorted). A \emph{prefix reversal} $\beta
=\beta(1,j)$ for some $3\leq j\leq n+1$ applied to $\pi$ reverses
the elements $\pi_1,\dots,\pi_{j-1}$ and thus transforms $\pi$ into
permutation
$\pi\cdot\beta=[\pi_0,\pi_{j-1},\ldots,\pi_1,\pi_j,\ldots,\pi_{n+1}]$.
A \emph{prefix transposition} $\tau=\tau(1,j,k)$ for some $2\leq
j\leq n$ and some $3\leq k\leq n+1$ such that $k\notin [1,j]$ cuts
the elements $\pi_1,\ldots,\pi_{j-1}$ and pastes between $\pi_{k-1}$
and $\pi_k$ and thus transforms $\pi$ into permutation
$\pi\cdot\tau=[\pi_0,\pi_j,\ldots,\pi_{k-1},\pi_1,\ldots,\pi_{j-1},\pi_k,\ldots,\pi_{n+1}]$.

An \emph{identity permutation $\iota_{n}$} is a permutation such
that $\pi_i=i$ for all $1\leq i\leq n$. Given two permutations, the
problem of sorting one permutation to another is equivalent to the
problem of sorting a given permutation to the identity permutation.
The \emph{prefix reversal and prefix transposition distance
$d(\pi)$} between $\pi$ and $\iota$ is the minimum number of
operations such that $\pi\cdot o_1\cdot o_2\cdot\ldots\cdot
o_{d(\pi)}=\iota$, where each operation $o_i$ is either a prefix
reversal $\beta$ or a prefix transposition $\tau$. The problem of
\emph{sorting by prefix reversals and prefix transpositions} is,
given a permutation $\pi$, to find a shortest sequence of prefix
reversals and prefix transpositions such that permutation $\pi$
transforms into the identity permutation $\iota$, i.e. finding the
distance $d(\pi)$.

A \emph{breakpoint} for this problem is a position $i$ of a permutation $\pi$
such that $|\pi_i-\pi_{i-1}|\neq 1$, and $2\leq i\leq n$. By definition,
position 1 (beginning of the permutation) is always considered a
breakpoint. Position $n + 1$ (end of the permutation) is considered a
breakpoint when $\pi_n\neq n$. We denote by $b(\pi)$ the number of
breakpoints of permutation $\pi$. Therefore, $b(\pi)\geq 1$ for any
permutation $\pi$ and the only permutations with exactly one
breakpoint are the identity permutations ($\pi=\iota_n$ , for all $n$).

The \emph{breakpoint graph} $G_\pi$ of $\pi$ is an undirected multi
graph whose vertices are $\pi_i$, for $0\leq i\leq n+1$, and edges
are of two types: \emph{grey} and \emph{black}. For each $1\leq
i\leq n+1$, the vertices $\pi_i$ and $\pi_{i-1}$ are joined by a
black edge iff there is a breakpoint between them, i.e., iff
$|\pi_i-\pi_{i-1}|\ne 1$. For $0\leq j<i\leq n+1$ and $j\neq i-1$,
there is a grey edge between $\pi_i$ and $\pi_j$ iff
$|\pi_i-\pi_{j}|=1$.

For convenience of illustration, in this paper the vertices of
$G_\pi$ are drawn horizontally from left to right in the order of
$\pi_0,\pi_1,\ldots,\pi_{n+1}$, the black edges
are drawn by horizontal lines, and the grey edges are drawn by
dotted arcs.
%---------------------------------------------------------------------
\section{3-approximation algorithm for prefix reversals and prefix transpositions}
\label{se:tr3}

\subsection{The lower bound}

For a permutation $\pi$ and an operation $o$, denote
$\triangle(\pi,o)=b(\pi)-b(\pi\cdot o)$ as the number of breakpoints
that are removed due to operation $o$.
Following are some important observations about breakpoints.

\begin{lemma}
\label{le:rev} $\triangle(\pi,\beta)\leq 1$.
\end{lemma}

\begin{lemma}
\label{le:tra} $\triangle(\pi,\tau)\leq 2$.
\end{lemma}

From Lemma~\ref{le:rev} and Lemma~\ref{le:tra} an optimal
algorithm for this problem can not remove more than two breakpoints
by a single operation. 
%As a breakpoint is present in the final permutation, a 
So, a lower bound follows.

\begin{theorem}
\label{th:lb2.67} $d(\pi)\geq \lfloor\frac{b(\pi)}{2}\rfloor$.
\end{theorem}

\subsection{The algorithm}
Our algorithm works on considering different orientations of grey
and black edges. Note that if a permutation is not sorted there must be at least two grey
edges in the breakpoint graph and each grey edge will be incident to two black edges.
A grey edge with its two adjacent black edges must be of one of the four types as shown in Fig.~\ref{fig:srtd}.

\iffalse
\begin{figure}
  \centering
  \includegraphics[width=.75\columnwidth]{4types}
  \caption{4 types of grey edges}
  \label{fig:rtd0}
\end{figure}
\fi

\begin{lemma}
\label{le:type4} 
%For a Type 1 edge with its grey edge having end points at $\pi_1$ and $\pi_j$, 
Let $(\pi_1,\pi_j)$ be a Type 1 grey edge. Then 
there exists at least one black edge $(\pi_{i-1},\pi_{i})$ for some $2\leq i\leq j$.
\end{lemma}

\begin{proof}
If no such black edge exists, then subsequence
$\pi_1\pi_2\ldots\pi_j$ is sorted. But in that case $(\pi_1,\pi_j)$
would not be a grey edge. 
\qed
\end{proof}

We call such a black edge a \emph{trapped black edge}. 

In our algorithm we scan the permutation from left to right to find the first black edge 
incident to a grey edge. There are four possible scenarios for the four types of edges. 
We consider the scenarios in the order as presented below in Fig.~\ref{fig:srtd} and apply 
a prefix transposition or a prefix reversal accordingly.

\begin{figure}
  \centering
  \input{srtd.pstex_t}
  \caption{Edge types and Scenarios of SortByRT3.} %for sorting by sorting by sorting by prefix reversals and prefix transpositions}
  \label{fig:srtd}
\end{figure}

\begin{lemma}
\label{le:gcases} 
Given  a permutation $\pi$ and its associated breakpoint graph $G(\pi)$, 
if any of the following two conditions is satisfied, 
then a prefix reversal or a prefix transposition can be applied to $\pi$ 
such that it removes at least one breakpoint.

\begin{enumerate} 
\item $G(\pi)$ contains a grey edge $(\pi_1,\pi_j)$ of Type 1 or Type 2 with $\pi_1\neq1$.
\item $G(\pi)$ contains a grey edge $(\pi_i,\pi_j)$ of Type 3 with $\pi_1=1$.
\end{enumerate}
\end{lemma}

\begin{proof}
If $\pi_1\neq1$ and there is a grey edge $(\pi_1,\pi_j)$ of Type 1, then Scenario 1 is 
applicable: according to Lemma~\ref{le:type4}, there exists a trapped black edge
$(\pi_{i-1},\pi_{i})$ for some $2\leq i\leq j$ and we apply a prefix transposition 
$\tau_1(1,i,{j+1})$ that creates adjacency between $\pi_{1}$ and $\pi_{j}$
without introducing any new breakpoint. 
If on the other hand the grey edge $(\pi_1,\pi_j)$ is of Type 2, then Scenario 2 is applicable:
apply a prefix reversal $\beta_2(1,j)$ that removes a breakpoint.

If $\pi_1=1$ and the first grey edge is $(\pi_i,\pi_j)$, then $\pi_{0},\pi_1\ldots\pi_{i}$ is sorted. 
If $(\pi_i,\pi_j)$ is of Type 3, then a prefix transposition $\tau_3(1,{i+1},j)$ removes one breakpoint 
according to Scenario 3. 
\qed
\end{proof}

\begin{lemma}
\label{le:bcase} Given a permutation $\pi$ and its associated breakpoint graph $G(\pi)$,
if none of the Scenario 1, 2 and 3 is applicable, then a prefix reversal can be applied
that does not remove any breakpoint but is followed by two subsequent operations removing 
at least two breakpoints.
\end{lemma}

\begin{proof}
If scenario 1 or 2 is not applicable, then $\pi_1=1$. 
Let $\pi_1,\pi_2\ldots\pi_{i-1}$, for some $1<i<n$, be the largest subsequence that is already sorted. 
Then there is a breakpoint between $\pi_{i-1}$ and $\pi_i$. 
If the grey edge adjacent to $\pi_{i-1}$ is not of Type 3, then it must be of Type 4.
%Its one endpoint is at $\pi_{i-1}$ and let 
Let the other endpoint of the grey edge be $\pi_{j-1}$. 
So, we can apply, according to Scenario 4, a prefix reversal $\beta_4(1,j)$ that does not remove any breakpoint but 
%$\beta_4(1,j)$ to transform the permutation into $\pi=\pi_0\pi_{j-1}\ldots\pi_i\pi_{i-1}\ldots\pi_1\pi_j\ldots\pi_n\pi_{n+1}$. 
%This causes the breakpoint to be at $(\pi_0,\pi_1)$ and 
causes the grey edge to become of Type 2. 
Then in the next step Scenario 2 will be applicable with a prefix reversal $\beta_2(j-1,i-1)$ that will remove one breakpoint. 
After applying $\beta_2$, $(\pi_0,\pi_i)$ will become a breakpoint with $i=1$ and $\pi_i\neq 1$.
%in the new permutation. 
Hence, again, either Scenario 1 or Scenario 2 will be applicable, which will further remove a breakpoint.
So, as a whole, we get two consecutive operations removing at least two
breakpoints after applying a reversal that does not remove a breakpoint.
\qed
\end{proof}

Our algorithm (SortByRT3) is summarized in Algorithm~1. It
clearly runs in polynomial time.

\begin{algorithm}
\label{al:a2.67app} \caption{SortByRT3($\pi$)} \label{Algorithm1}
\begin{algorithmic}
\STATE Construct breakpoint graph $G_\pi$ of $\pi$
\WHILE{there is a breakpoint}
\IF{$\pi_1\neq 1$ }
%    \STATE Find the position $(\pi_{i-1},\pi_i)$ of second
%    breakpoint
%\ELSE
%    \STATE set $i=1$
% \ENDIF
       \IF{Scenario 1 is applicable} 
       \STATE apply a prefix transposition $\tau_1$
       \ELSIF{Scenario 2 is applicable} 
       \STATE apply a prefix reversal $\beta_2$
       \ENDIF
\ELSE
       \IF {Scenario 3 is applicable} 
       \STATE apply a prefix transposition $\tau_3$
       \ELSE  
       \STATE apply a prefix reversal $\beta_4$
       \ENDIF
\ENDIF
\ENDWHILE
\end{algorithmic}
\end{algorithm}

\begin{theorem}
\label{th:sortbyrt3}
SortByRT3 is a 3-approximation algorithm.
\end{theorem}

\begin{proof}
By Lemma~\ref{le:gcases}, if any of the Scenario 1, 2 or 3 is applicable, 
then the algorithm can remove at least one breakpoint at each step. 
Otherwise according to Lemma~\ref{le:bcase} it removes at least two breakpoints in three steps. 
Hence, it sorts $\pi$ in
at most $\frac{3(b(\pi)-1)}{2}$ operations. 
By Theorem~\ref{th:lb2.67}, $d(\pi)\ge \lfloor\frac{b(\pi)}{2}\rfloor$.
So, we get an approximation ratio of $\rho \leq3$. \qed
\end{proof}

%----------------------------------------------------------------
\section{2-approximation algorithm}
\label{se:tr2}

Now we improve the ratio considering a third  rearrangement operation,
called \emph{prefix transreversal}.
A \emph{prefix transreversal} $\beta\tau=\beta\tau(1,j,k)$ for some
$2\leq j\leq n$ and some $3\leq k\leq n+1$ such that  $k\notin
[1,j]$ reverses the elements $\pi_1,\ldots,\pi_{j-1}$ and then
pastes it between $\pi_{k-1}$ and $\pi_k$ and thus transforms $\pi$ into permutation
$\pi\cdot\beta\tau=[\pi_0,\pi_j,\ldots,\pi_{k-1},\pi_{j-1},\ldots,\pi_{1},\pi_k,\ldots,\pi_{n+1}]$.

\subsection{The lower bound}

Another important observation about breakpoints regarding prefix
transreversals is the following.

\begin{lemma}
\label{le:trarev} $\triangle(\pi,\beta\tau)\leq 2$.
\end{lemma}

From Lemma~\ref{le:rev}, Lemma~\ref{le:tra} and
Lemma~\ref{le:trarev} a lower bound for sorting by prefix reversals and prefix transreversals
is the following.

\begin{theorem}
\label{th:lb2} $d(\pi)\ge \lfloor\frac{b(\pi)}{2}\rfloor$.
\end{theorem}

\subsection{The algorithm}
The next lemma is the key to our 2-approximation.

\begin{lemma}
\label{le:steps4} 
Let $\pi$ be a permutation with $\pi_1=1$ and let its associated breakpoint graph be $G(\pi)$.
If $G(\pi)$ contains a grey edge of Type 4, then a prefix transreversal can be applied
that removes at least one breakpoint.
\end{lemma}

\begin{proof}
Let the Type 4 grey edge be $(\pi_{i-1},\pi_{j-1})$ with its two adjacent black edges 
$(\pi_{i-1},\pi_i)$ and $(\pi_{j-1},\pi_j)$. 
We can apply a prefix transreversal $\beta\tau(1,i,j)$ creating an adjacency between 
$\pi_{i-1}$ and $\pi_{j-1}$ and thus removing a breakpoint.
\qed
\end{proof}

The above lemma along with Lemma~\ref{le:gcases} proves that in every situation 
at least one breakpoint is removed by each operation.

\begin{lemma}
\label{le:ub2} For every permutation $\pi$, we have $d({\pi})\leq
b({\pi})-1$.
\end{lemma}

\begin{theorem}
\label{th:range} 
$\lfloor\frac{b(\pi)}{2}\rfloor \leq d({\pi}) \leq b({\pi})-1$.
\end{theorem}

\begin{theorem}
\label{th:sortbyrt2} 
An algorithm (let us call it SortByRT2) that produces prefix reversals, prefix transpositions,
and/or prefix transreversals according to Lemma~\ref{le:steps4} is an approximation algorithm with factor 2
for sorting by prefix reversals and prefix transreversals.
\end{theorem}

%--------------------------------------------------------------------
%********************************************************************
\section{3-approximation algorithm with forward march}
\label{se:march}

In this section we introduce a new concept that we call \emph{forward march}.
At the very beginning or after applying a prefix reversal or a
prefix transposition, a prefix $\pi_0,\ldots,\pi_i$, for $0\leq i\leq
n+1$ may be already sorted.  In this case we update $\pi$ as the unsorted suffix 
of $\pi$, i.e., as $\pi=\pi_{i},\ldots,\pi_{n+1}$ and the size of $\pi$ 
is reduced by $i$, i.e., the value of $n$ is updated as $n=n-i$.
The next prefix reversal or prefix transposition is applied on
updated $\pi$. This concept of moving
forward along with the  sorting is called \emph{forward march}.

For our algorithm with forward march we redefine breakpoint and breakpoint
graph. In the redefined \emph{breakpoint graph} $G_\pi$ of $\pi$ there is a black
edge between $\pi_i$ and $\pi_{i+1}$ iff there is a breakpoint between them, i.e., iff
$|\pi_i-\pi_{i+1}|\ne 1$. Clearly, $\pi$ is sorted iff it has no
breakpoint. Note that at any time $\pi_0$ is the last element in the sorted part
of the permutation and there always exists a black edge between
$\pi_0$ and $\pi_1$. We call this black edge the \emph{starting
black edge}.

\subsection{The lower bound}

Due to breakpoint redefinition some of our previous observations are
modified.

\begin{lemma}
\label{le:rev3} 
$\triangle(\pi,\beta)\leq 2$.
\end{lemma}

\begin{lemma}
\label{le:tra3} 
$\triangle(\pi,\tau)\leq 3$.
\end{lemma}

 From Lemma~\ref{le:rev3} and Lemma~\ref{le:tra3} new lower bound is the following.

\begin{theorem}
\label{th:lb3} $d(\pi)\geq \frac{b(\pi)}{3}$.
\end{theorem}

\subsection{The algorithm}
Our algorithm works on considering different orientations of grey
and black edges. If $\pi$ is unsorted then it has at least two grey edges
and at least one black edge in addition to the starting black edge
$(\pi_0,\pi_1)$.

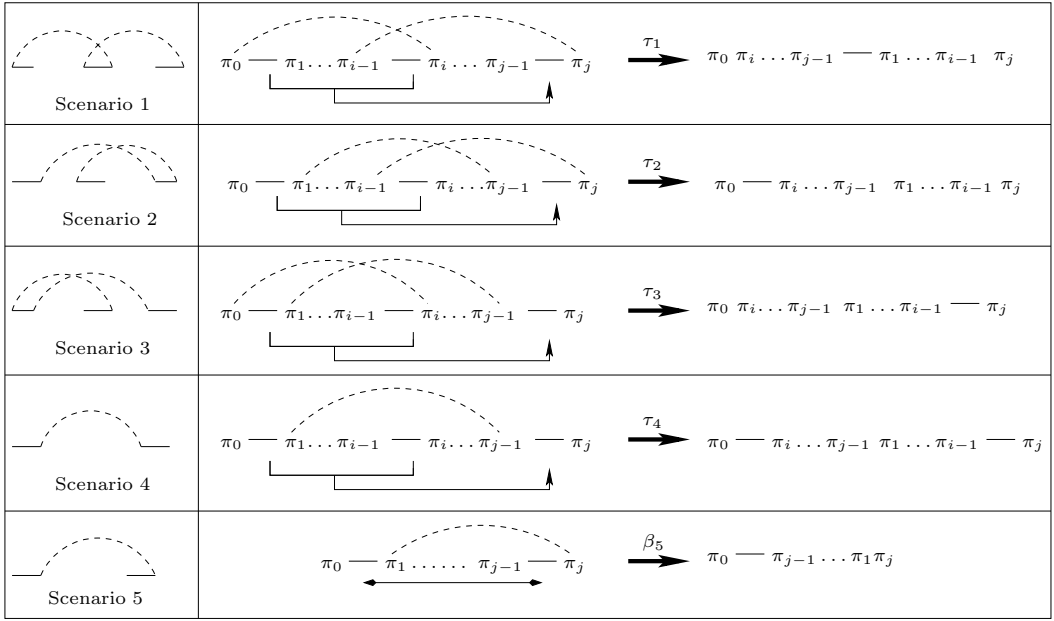
\begin{figure}[th]
  \centering
  \input{rtd.pstex_t}
  \caption{Scenarios of SortByRTwFM3}
  \label{fig:rtd}
\end{figure}

We consider different orientations of the four edge types of Fig.~\ref{fig:srtd}
described in Section~\ref{se:tr3}. 
We try five scenarios in the order shown in Fig.~\ref{fig:rtd}, apply 
a prefix transposition or a prefix reversal accordingly and perform a forward march
if possible. In fact, Lemma~\ref{le:ends} proves that Scenario 4 and 5 are sufficient to
sort $\pi$, and Scenario 1, 2 and 3 improve practical performance of our algorithm without
affecting approximation ratio. Our algorithm (SortByRTwFM3) is summarized in Algorithm~2. 
It clearly runs in polynomial time.

The following lemma is immediate from the scenarios presented in Fig.~\ref{fig:rtd}.

\begin{lemma}
\label{le:ub3} After each prefix reversal or prefix transposition
the number of breakpoints is reduced by at least one.
\end{lemma}

The following lemma proves that our algorithm always terminates by sorting the permutation.

\begin{lemma}
\label{le:ends} Scenario 4 and Scenario 5 are sufficient to sort the
permutation.
\end{lemma}

\begin{proof}
Since $\pi_1\neq 1$, we always have a grey edge of Type 1 or Type 2 
whose left black edge is $(\pi_o,\pi_1)$. 
%the starting black edge. 
If the grey edge is of Type 1, then by Lemma~\ref{le:type4} we can find a
trapped black edge and can apply Scenario 4. On the other hand, if the
grey edge is of Type 2, then Scenario 5 is applicable. By
Lemma~\ref{le:ub3} since every scenario reduces at least one
breakpoint, Scenario 4 and Scenario 5 can successfully sort the
permutation. \qed
\end{proof}

\begin{algorithm}
\label{al:a3app} \caption{SortByRTwFM3($\pi$)} \label{Algorithm2}
\begin{algorithmic}
\STATE Construct breakpoint graph $G_\pi$ of $\pi$ \WHILE{there is a
black edge}
  \IF{Scenario 1 is found}
    \STATE apply a prefix transposition $\tau_1$
  \ELSIF{Scenario 2 is found}
    \STATE apply a prefix transposition $\tau_2$
  \ELSIF {Scenario 3 is found}
    \STATE apply a prefix transposition $\tau_3$
  \ELSIF {Scenario 4 is found}
    \STATE apply a prefix transposition $\tau_4$
  \ELSE
    \STATE apply a prefix reversal $\beta_5$
  \ENDIF
  \STATE $G_\pi\leftarrow G_\pi$ after applying the operation
  \STATE update starting black edge to the first black edge of $G_\pi$ \hspace{25pt} // Forward march
\ENDWHILE
\end{algorithmic}
\end{algorithm}

\begin{theorem}
\label{th:sortbyrtwfm3}
SortByRTwFM3 is a 3-approximation algorithm.
\end{theorem}

\begin{proof}
By Lemma~\ref{le:ends} our algorithm successfully sorts a given
permutation $\pi$ and by Lemma~\ref{le:ub3} it sorts $\pi$ in at
most $b(\pi)$ operations. By Lemma~\ref{th:lb3},
$d(\pi)\ge\frac{b(\pi)}{3}$. So, we get an approximation ratio of
$\rho \leq3$. \qed
\end{proof}

\section{Adaptive approximation ratios}
\label{se:alt}

The algorithms presented in this paper realize their worst case approximation ratios 
as a result of combination of the best
case behavior of an optimal algorithm, where no prefix reversal is
applied, and a worst case behavior of our algorithm, where no prefix
transposition (or prefix transreversal) is applied. 
This is due to the inferiority of prefix reversals to prefix transpositions with respect
to their ability to remove breakpoints.
However, it is expected that an optimal algorithm would apply both operations.
Motivated by the above observation, we derive adaptive approximation ratios for our 
algorithms in terms of the number of prefix reversals, $r$, applied by an optimal algorithm.
Although there is no change in the upper bound of the algorithm, the
approximation ratio will improve, because of the increased lower
bound.

%+++++++++++++++++++++++++++++++++++++++++++++++++++++++++++++++++

\begin{theorem}
\label{le:ptr2.67} 
When $r$ prefix reversals are applied by an optimal
algorithm, SortByRT3 sorts a permutation $\pi$ with
approximation ratio $\rho_r\le3-\frac{3r}{b(\pi)+r-1}$.
\end{theorem}

\begin{proof}
For $b(\pi)\geq r$, if an optimal algorithm uses $r$ prefix
reversals, it removes at most $r$ breakpoints (by Lemma
\ref{le:rev}). For the remaining $(b(\pi)-1)-r$ breakpoints it must
use at least $\lfloor\frac{b(\pi)-r}{2}\rfloor$ operations (by
Theorem~\ref{th:lb2.67}). 

As a result, $d(\pi)\geq
r+\lfloor\frac{b(\pi)-r}{2}\rfloor\geq
r+\frac{b(\pi)-1-r}{2} = \frac{b(\pi)-1}{2}+\frac{r}{2}$

Therefore,
$\rho_r\leq\frac{\frac{3(b(\pi)-1)}{2}}{\frac{b(\pi)+r-1}{2}}
            =\frac{3(b(\pi)-1)}{b(\pi)+r-1}
             = 3-\frac{3r}{b(\pi)+r-1}$
\qed
\end{proof}
%+++++++++++++++++++++++++++++++++++++++++

\begin{theorem}
\label{le:ptr2}
When $r$ prefix reversals are applied by an optimal
 algorithm, SortByRT2 sorts a permutation $\pi$ with
approximation ratio $\rho_r\le2-\frac{2r}{b(\pi)+r-1}$.
\end{theorem}

\begin{proof}
An optimal algorithm removes at most $r$ breakpoints by $r$ prefix
reversals (by Lemma \ref{le:rev}). For the remaining $(b(\pi)-1-r)$ breakpoints it must
use at least $\lfloor\frac{b(\pi)-r}{2}\rfloor$ operations (by
Theorem~\ref{th:lb2}). 

As a result, $d(\pi)\geq
r+\lfloor\frac{b(\pi)-r}{2}\rfloor\geq
r+\frac{b(\pi)-1-r}{2} = \frac{b(\pi)-1}{2}+\frac{r}{2}$

Therefore,
$\rho_r\leq\frac{b(\pi)-1}{\frac{b(\pi)+r-1}{2}}
            =\frac{2b(\pi)-2+2r-2r}{b(\pi)+r-1}
             = 2-\frac{2r}{b(\pi)+r-1}$
\qed
\end{proof}

%-------------------------------------------------------------------

\begin{theorem}
\label{th:fm3} 
When $r$ prefix reversals are performed by an optimal algorithm, 
SortByRTwFM3 sorts a permutation $\pi$ with approximation ratio $\rho_r\le3-\frac{3r}{b(\pi)+r}$.
\end{theorem}

\begin{proof}
For $b(\pi)\geq2r$ an optimal algorithm applies $r$ prefix reversals 
to remove at most $2r$ breakpoints (by Lemma~\ref{le:rev3}).
For the remaining $b(\pi)-2r$ breakpoints it must use
at least $\frac{b{\pi}-2r}{3}$ operations (by Theorem~\ref{th:lb3}).

Therefore, $d(\pi)\geq r+\frac{b(\pi)-2r}{3} =
\frac{b(\pi)}{3}+\frac{r}{3}$

So,
$\rho_r\leq\frac{b(\pi)}{\frac{b(\pi)+r}{3}}
            =\frac{3b(\pi)+3r-3r}{b(\pi)+r}
             = 3-\frac{3r}{b(\pi)+r}$
\qed
\end{proof}

\section{Experimental Results}
\label{se:exp}

\begin{figure}[htbp]
\begin{center}
\subfigure[SortByRT3]{\includegraphics[width=.45\textwidth]{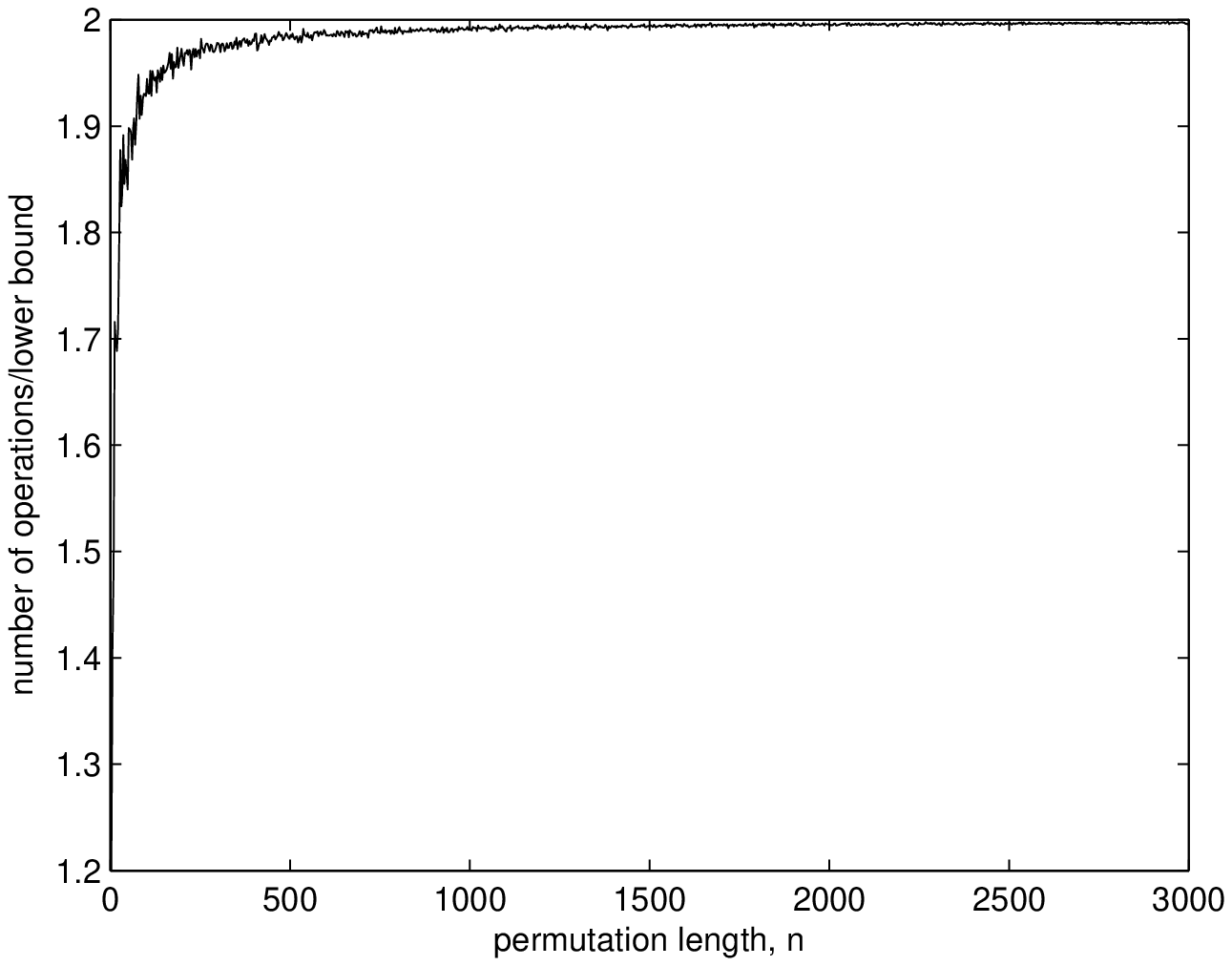}}
\subfigure[Adaptive SortByRT3]{\includegraphics[width=.45\textwidth]{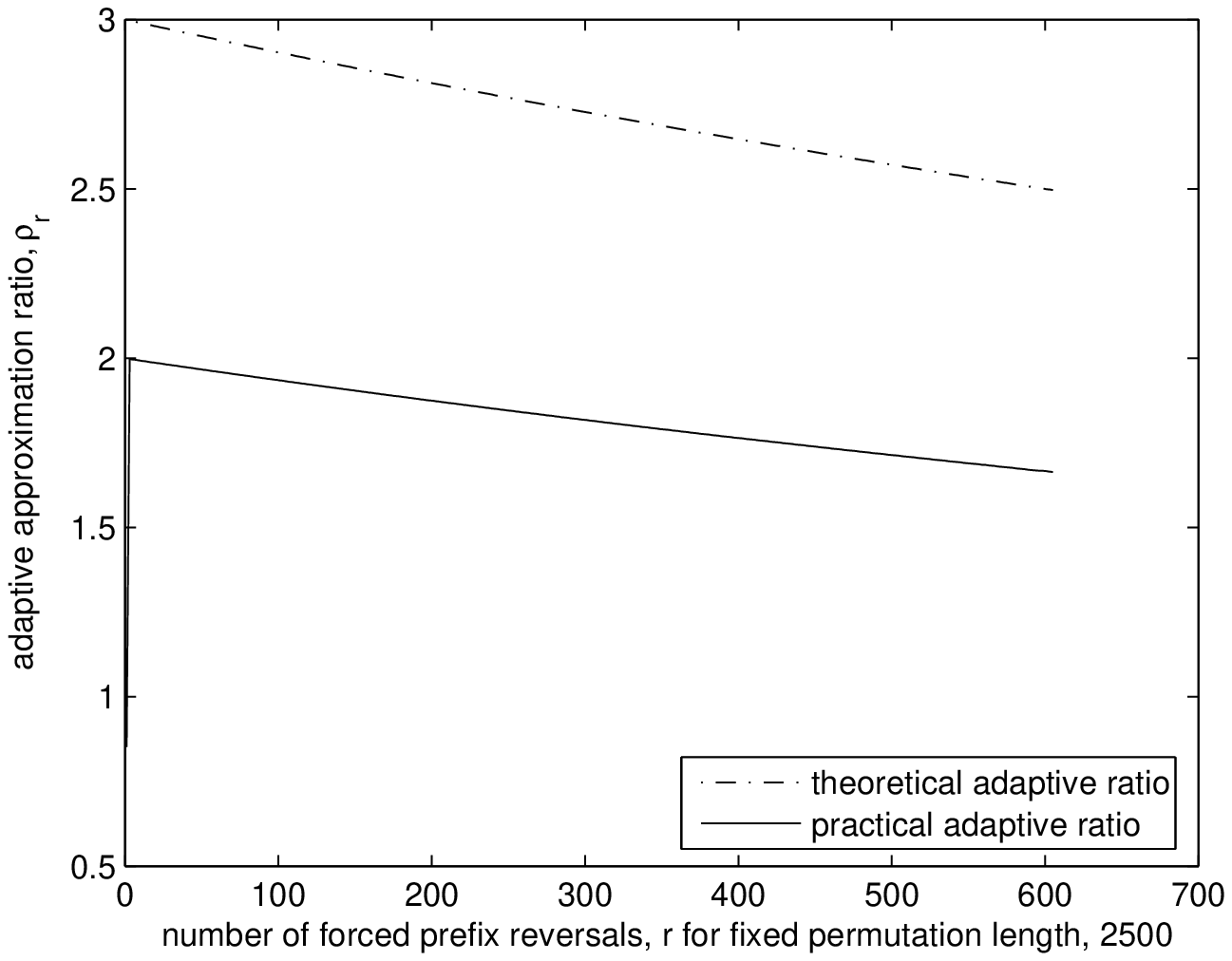}}
\subfigure[SortByRT2]{\includegraphics[width=.45\textwidth]{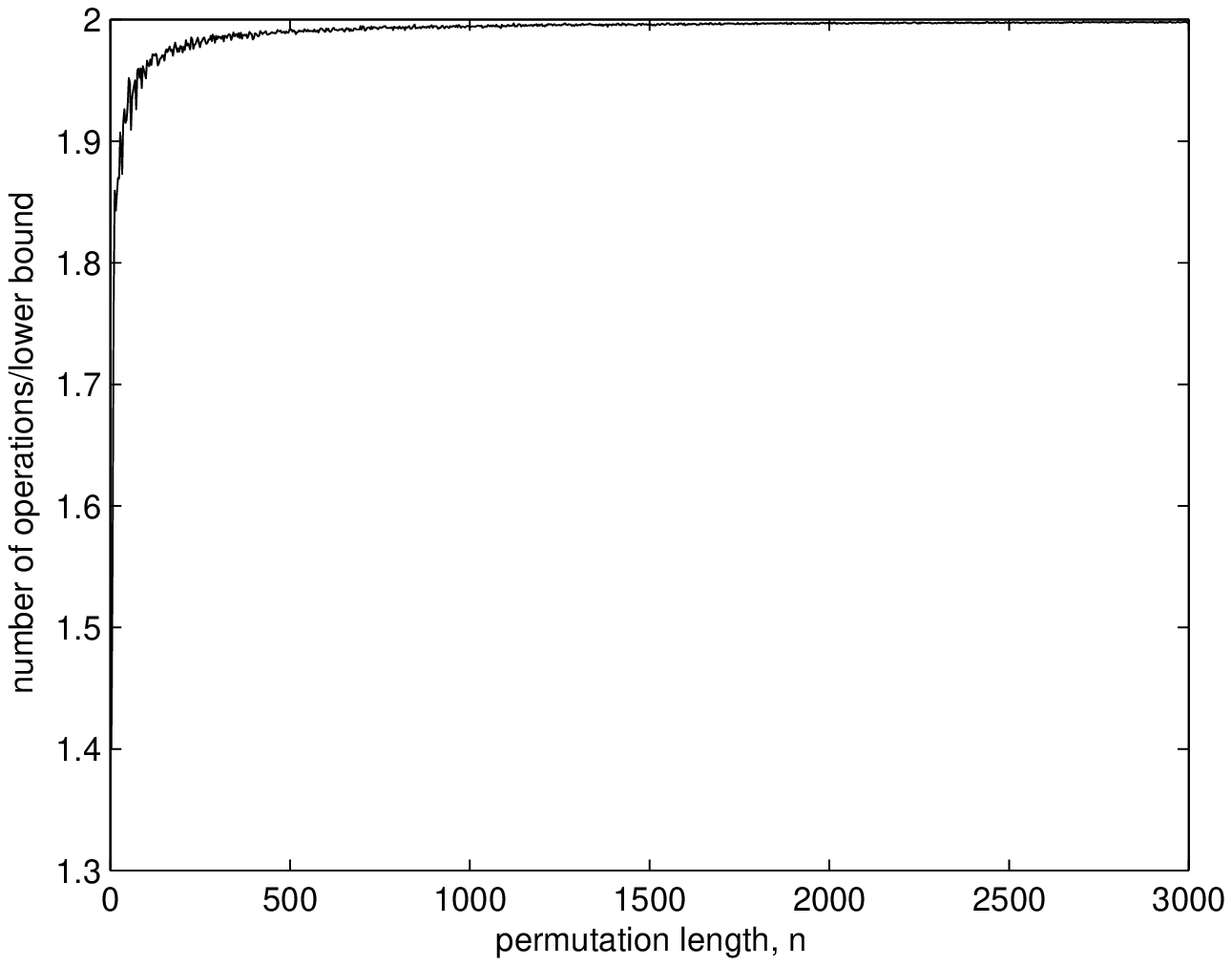}}
\subfigure[Adaptive SortByRT2]{\includegraphics[width=.45\textwidth]{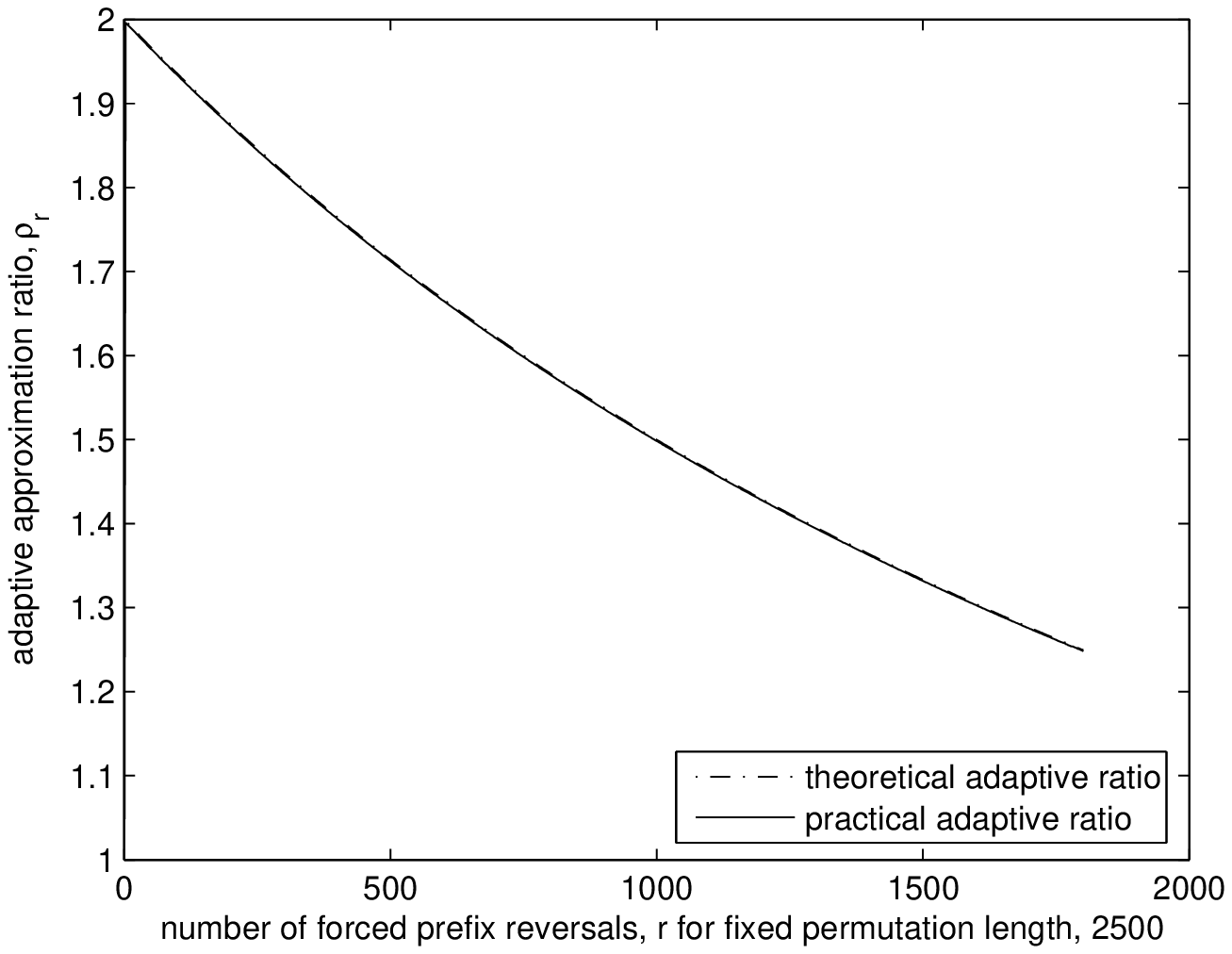}}
\subfigure[SortByRTwFM3]{\includegraphics[width=.45\textwidth]{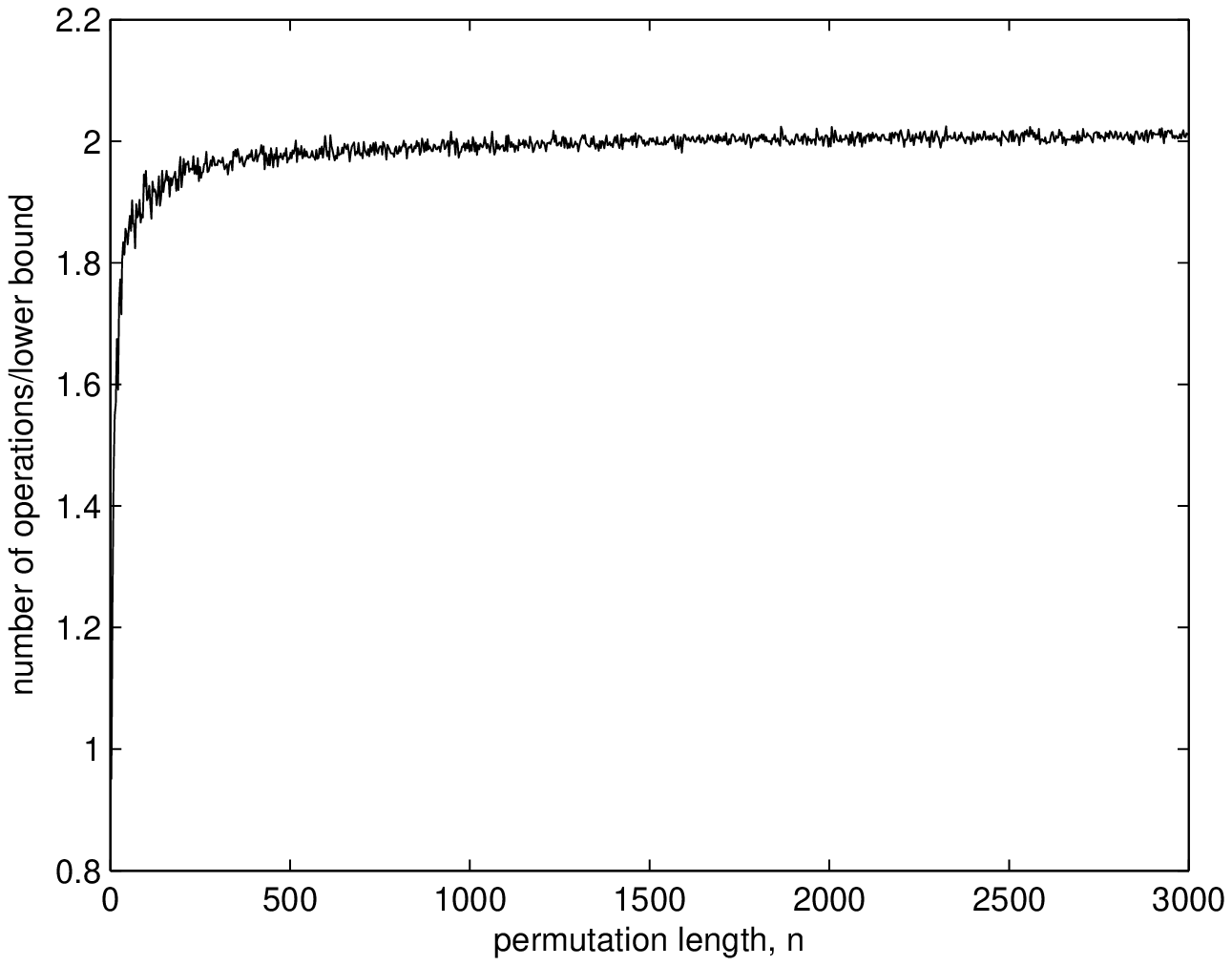}}
\subfigure[Adaptive SortByRTwFM3]{\includegraphics[width=.45\textwidth]{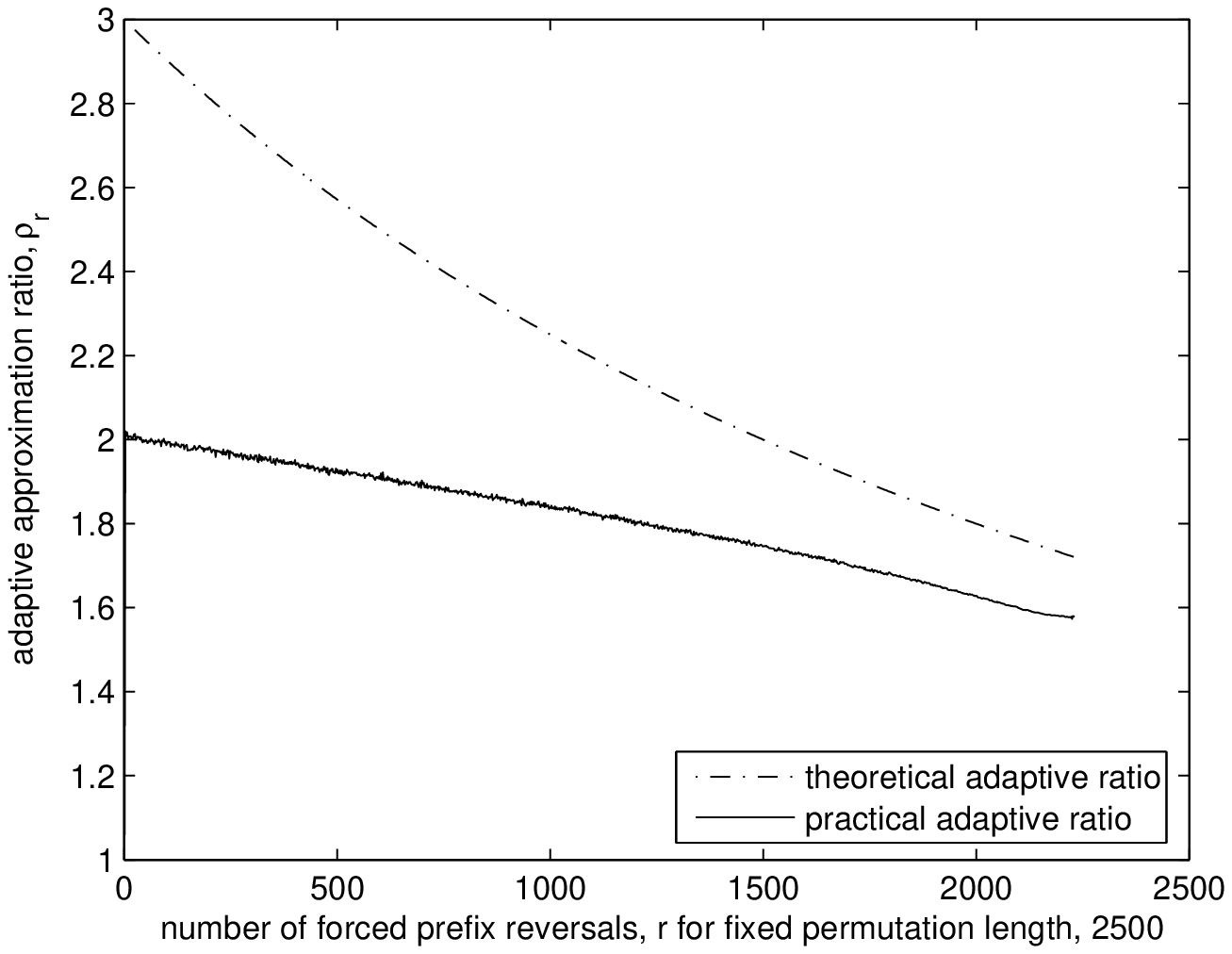}}
\caption{Experimental results} 
\label{fig:experiment}
\end{center}
\end{figure}

We have implemented our algorithms and tested their average performance 
on above 60,000 permutations taken randomly of size up to 3000. 
In each case the cost of solution given by proposed algorithm is compared 
to corresponding lower bound instead of comparing with the cost of optimal solution. 
For both SortByRT3 (Theorem~\ref{th:sortbyrt3}) and SortByRTwFM3 (Theorem~\ref{th:sortbyrtwfm3}) 
worst cases occur very few times in practice and shows ratio near 2 in average 
(Fig.~\ref{fig:experiment} (a) and (e) respectively).
For SortByRT2 (Theorem~\ref{th:sortbyrt2}) the practical ratio is no better than the 
theoretical one due to its strategy of choosing scenarios most of the time which remove 
only $1$ breakpoint in one operation (Fig.~\ref{fig:experiment}(c)).

The ultimate effect of inducing inferior operations 
(Theorem~\ref{le:ptr2.67},~\ref{le:ptr2} and~\ref{th:fm3})
is the increase of lower bounds and thereby the decrease of approximation ratios.
This is reflected in Fig.~\ref{fig:experiment}(b), (d) and (f), respectively,
as the corresponding theoretical and practical curves decrease and become closer
to the optimal with the increase of $r$.

Observe that if we could compare our algorithms with corresponding optimal algorithms, 
then the experimental ratios would be better. 
%Experimental results also show that as the number of prefix reversals applied by an optimal algorithm increases, performance
%of the algorithms improves steadily.

\iffalse**********************************
\begin{figure}[ht]
\begin{center}
\subfigure[3-approximation
algorithm]{\includegraphics[width=.49\textwidth]{3r1_2}}
\subfigure[Forced prefix
reversals]{\includegraphics[width=.49\textwidth]{rr2500_2}}
\caption{Experimental results} \label{fig:exp}
\end{center}
\end{figure}
*******************************************\fi

\section{Conclusion}
\label{se:con}
In this paper we have studied some variations of the pancake flipping problem from the view point of sorting
 unsigned permutations.
We have given a 3-approximation algorithm for sorting by prefix reversals and prefix transpositions. 
Then we considered a third operation, called prefix transreversal, and provided a 3-approximation algorithm. 
We also introduced a new concept called 
forward march where we skip over the sorted prefix of the permutation and apply
operations on some prefix of the unsorted suffix of the permutation and contributed
a 3-approximation algorithm. 
% for sorting by prefix reversals and prefix transpositions with forward march. 
We have further analyzed the problems in more practical way and 
presented better approximation ratios when a certain number of
inferior operations (i.e., prefix reversals) are applied by an optimal algorithm.
An experimental study shows that our algorithms performs much better in practice than 
suggested by their approximation ratios. 

It will be interesting to redefine the problems to ``force'' certain number of inferior operations 
and analyze the approximation ratios of the algorithms. This idea can be applied for other 
combination of more than one operation. The complexity of the problems are also unknown. In future it would
be interesting to see whether the problems are NP-hard.

\bibliography{forward_current.bib}

\end{document}

%% file: srtd.pstex_t
\begin{picture}(0,0)%
\includegraphics{srtd.pstex}%
\end{picture}%
\setlength{\unitlength}{2368sp}%
\begingroup\makeatletter\ifx\SetFigFont\undefined%
\gdef\SetFigFont#1#2#3#4#5{%
  \reset@font\fontsize{#1}{#2pt}%
  \fontfamily{#3}\fontseries{#4}\fontshape{#5}%
  \selectfont}%
\fi\endgroup%
\begin{picture}(11349,5349)(589,-4198)
\put(7351,164){\makebox(0,0)[lb]{\smash{{\SetFigFont{7}{8.4}{\rmdefault}{\mddefault}{\updefault}$\tau_1$}}}}
\put(3376,-2536){\makebox(0,0)[lb]{\smash{{\SetFigFont{7}{8.4}{\rmdefault}{\mddefault}{\updefault}$\pi_0-\pi_1\ldots\pi_{i}-\pi_{i+1}\ldots\pi_{j-1}-\pi_j$}}}}
\put(3376,-61){\makebox(0,0)[lb]{\smash{{\SetFigFont{7}{8.4}{\rmdefault}{\mddefault}{\updefault}$\pi_0-\pi_1\ldots\pi_{i-1}-\pi_i\ldots\pi_{j}-\pi_{j+1}$}}}}
\put(3376,-3736){\makebox(0,0)[lb]{\smash{{\SetFigFont{7}{8.4}{\rmdefault}{\mddefault}{\updefault}$\pi_0-\pi_1\ldots\pi_{i-1}-\pi_i\ldots\pi_{j-1}-\pi_j$}}}}
\put(8326,-61){\makebox(0,0)[lb]{\smash{{\SetFigFont{7}{8.4}{\rmdefault}{\mddefault}{\updefault}$\pi_0-\pi_i\ldots\pi_{j}\pi_1\ldots\pi_{i-1}-\pi_{j+1}$}}}}
\put(8326,-1336){\makebox(0,0)[lb]{\smash{{\SetFigFont{7}{8.4}{\rmdefault}{\mddefault}{\updefault}$\pi_0-\pi_{j-1}\ldots\pi_1\pi_j$}}}}
\put(8326,-2536){\makebox(0,0)[lb]{\smash{{\SetFigFont{7}{8.4}{\rmdefault}{\mddefault}{\updefault}$\pi_0-\pi_{i+1}\ldots\pi_{j-1}-\pi_1\ldots\pi_{i}\pi_j$}}}}
\put(7351,-1036){\makebox(0,0)[lb]{\smash{{\SetFigFont{7}{8.4}{\rmdefault}{\mddefault}{\updefault}$\beta_2$}}}}
\put(7351,-2311){\makebox(0,0)[lb]{\smash{{\SetFigFont{7}{8.4}{\rmdefault}{\mddefault}{\updefault}$\tau_3$}}}}
\put(3976,-1336){\makebox(0,0)[lb]{\smash{{\SetFigFont{7}{8.4}{\rmdefault}{\mddefault}{\updefault}$\pi_0-\pi_1\ldots\pi_{j-1}-\pi_j$}}}}
\put(8326,-3736){\makebox(0,0)[lb]{\smash{{\SetFigFont{7}{8.4}{\rmdefault}{\mddefault}{\updefault}$\pi_0-\pi_{j-1}\ldots\pi_{i}-\pi_{i-1}\ldots\pi_1-\pi_j$}}}}
\put(7351,-3511){\makebox(0,0)[lb]{\smash{{\SetFigFont{7}{8.4}{\rmdefault}{\mddefault}{\updefault}$\beta_4$}}}}
\end{picture}%

%% file: rtd.pstex_t
\begin{picture}(0,0)%
\includegraphics{rtd.pstex}%
\end{picture}%
\setlength{\unitlength}{2368sp}%
\begingroup\makeatletter\ifx\SetFigFont\undefined%
\gdef\SetFigFont#1#2#3#4#5{%
  \reset@font\fontsize{#1}{#2pt}%
  \fontfamily{#3}\fontseries{#4}\fontshape{#5}%
  \selectfont}%
\fi\endgroup%
\begin{picture}(10974,6474)(64,-5848)
\put(6751,-5086){\makebox(0,0)[lb]{\smash{{\SetFigFont{7}{8.4}{\rmdefault}{\mddefault}{\updefault}$\beta_5$}}}}
\put(7426, 14){\makebox(0,0)[lb]{\smash{{\SetFigFont{7}{8.4}{\rmdefault}{\mddefault}{\updefault}$\pi_0$}}}}
\put(7726, 14){\makebox(0,0)[lb]{\smash{{\SetFigFont{7}{8.4}{\rmdefault}{\mddefault}{\updefault}$\pi_i\ldots\pi_{j-1}$}}}}
\put(9226, 14){\makebox(0,0)[lb]{\smash{{\SetFigFont{7}{8.4}{\rmdefault}{\mddefault}{\updefault}$\pi_1\ldots\pi_{i-1}$}}}}
\put(10426, 14){\makebox(0,0)[lb]{\smash{{\SetFigFont{7}{8.4}{\rmdefault}{\mddefault}{\updefault}$\pi_j$}}}}
\put(7501,-1336){\makebox(0,0)[lb]{\smash{{\SetFigFont{7}{8.4}{\rmdefault}{\mddefault}{\updefault}$\pi_0$}}}}
\put(8176,-1336){\makebox(0,0)[lb]{\smash{{\SetFigFont{7}{8.4}{\rmdefault}{\mddefault}{\updefault}$\pi_i\ldots\pi_{j-1}$}}}}
\put(9376,-1336){\makebox(0,0)[lb]{\smash{{\SetFigFont{7}{8.4}{\rmdefault}{\mddefault}{\updefault}$\pi_1\ldots\pi_{i-1}$  $\pi_j$}}}}
\put(7426,-2611){\makebox(0,0)[lb]{\smash{{\SetFigFont{7}{8.4}{\rmdefault}{\mddefault}{\updefault}$\pi_0$   $\pi_i$$\ldots\pi_{j-1}$}}}}
\put(10351,-2611){\makebox(0,0)[lb]{\smash{{\SetFigFont{7}{8.4}{\rmdefault}{\mddefault}{\updefault}$\pi_j$}}}}
\put(8851,-2611){\makebox(0,0)[lb]{\smash{{\SetFigFont{7}{8.4}{\rmdefault}{\mddefault}{\updefault}$\pi_1\ldots\pi_{i-1}$}}}}
\put(8101,-4036){\makebox(0,0)[lb]{\smash{{\SetFigFont{7}{8.4}{\rmdefault}{\mddefault}{\updefault}$\pi_i\ldots\pi_{j-1}$}}}}
\put(7426,-4036){\makebox(0,0)[lb]{\smash{{\SetFigFont{7}{8.4}{\rmdefault}{\mddefault}{\updefault}$\pi_0$}}}}
\put(10726,-4036){\makebox(0,0)[lb]{\smash{{\SetFigFont{7}{8.4}{\rmdefault}{\mddefault}{\updefault}$\pi_j$}}}}
\put(9226,-4036){\makebox(0,0)[lb]{\smash{{\SetFigFont{7}{8.4}{\rmdefault}{\mddefault}{\updefault}$\pi_1\ldots\pi_{i-1}$}}}}
\put(3376,-5311){\makebox(0,0)[lb]{\smash{{\SetFigFont{7}{8.4}{\rmdefault}{\mddefault}{\updefault}$\pi_0$}}}}
\put(4951,-5311){\makebox(0,0)[lb]{\smash{{\SetFigFont{7}{8.4}{\rmdefault}{\mddefault}{\updefault} $\pi_{j-1}$              }}}}
\put(4051,-5311){\makebox(0,0)[lb]{\smash{{\SetFigFont{7}{8.4}{\rmdefault}{\mddefault}{\updefault}$\pi_1\ldots\ldots$}}}}
\put(5926,-5311){\makebox(0,0)[lb]{\smash{{\SetFigFont{7}{8.4}{\rmdefault}{\mddefault}{\updefault}$\pi_j$}}}}
\put(7426,-5236){\makebox(0,0)[lb]{\smash{{\SetFigFont{7}{8.4}{\rmdefault}{\mddefault}{\updefault}$\pi_0$}}}}
\put(8101,-5236){\makebox(0,0)[lb]{\smash{{\SetFigFont{7}{8.4}{\rmdefault}{\mddefault}{\updefault}$\pi_{j-1}\ldots\pi_1\pi_j$}}}}
\put(2326,-61){\makebox(0,0)[lb]{\smash{{\SetFigFont{7}{8.4}{\rmdefault}{\mddefault}{\updefault}$\pi_0$}}}}
\put(3001,-61){\makebox(0,0)[lb]{\smash{{\SetFigFont{7}{8.4}{\rmdefault}{\mddefault}{\updefault}$\pi_1$$\ldots\pi_{i-1}$}}}}
\put(4501,-61){\makebox(0,0)[lb]{\smash{{\SetFigFont{7}{8.4}{\rmdefault}{\mddefault}{\updefault}$\pi_i\ldots$ $\pi_{j-1}$        }}}}
\put(6001,-61){\makebox(0,0)[lb]{\smash{{\SetFigFont{7}{8.4}{\rmdefault}{\mddefault}{\updefault}$\pi_j$}}}}
\put(601,-511){\makebox(0,0)[lb]{\smash{{\SetFigFont{7}{8.4}{\rmdefault}{\mddefault}{\updefault}Scenario 1}}}}
\put(2401,-1336){\makebox(0,0)[lb]{\smash{{\SetFigFont{7}{8.4}{\rmdefault}{\mddefault}{\updefault}$\pi_0$}}}}
\put(3076,-1336){\makebox(0,0)[lb]{\smash{{\SetFigFont{7}{8.4}{\rmdefault}{\mddefault}{\updefault}$\pi_1$$\ldots\pi_{i-1}$}}}}
\put(6076,-1336){\makebox(0,0)[lb]{\smash{{\SetFigFont{7}{8.4}{\rmdefault}{\mddefault}{\updefault}$\pi_j$}}}}
\put(4576,-1336){\makebox(0,0)[lb]{\smash{{\SetFigFont{7}{8.4}{\rmdefault}{\mddefault}{\updefault}$\pi_i\ldots$$\pi_{j-1}$            }}}}
\put(676,-1711){\makebox(0,0)[lb]{\smash{{\SetFigFont{7}{8.4}{\rmdefault}{\mddefault}{\updefault}Scenario 2}}}}
\put(2326,-2686){\makebox(0,0)[lb]{\smash{{\SetFigFont{7}{8.4}{\rmdefault}{\mddefault}{\updefault}$\pi_0$}}}}
\put(5926,-2686){\makebox(0,0)[lb]{\smash{{\SetFigFont{7}{8.4}{\rmdefault}{\mddefault}{\updefault}$\pi_j$}}}}
\put(4426,-2686){\makebox(0,0)[lb]{\smash{{\SetFigFont{7}{8.4}{\rmdefault}{\mddefault}{\updefault}$\pi_i$$\ldots\pi_{j-1}$}}}}
\put(3001,-2686){\makebox(0,0)[lb]{\smash{{\SetFigFont{7}{8.4}{\rmdefault}{\mddefault}{\updefault}$\pi_1$$\ldots$$\pi_{i-1}$}}}}
\put(601,-3061){\makebox(0,0)[lb]{\smash{{\SetFigFont{7}{8.4}{\rmdefault}{\mddefault}{\updefault}Scenario 3}}}}
\put(2326,-4036){\makebox(0,0)[lb]{\smash{{\SetFigFont{7}{8.4}{\rmdefault}{\mddefault}{\updefault}$\pi_0$}}}}
\put(6001,-4036){\makebox(0,0)[lb]{\smash{{\SetFigFont{7}{8.4}{\rmdefault}{\mddefault}{\updefault}$\pi_j$}}}}
\put(4501,-4036){\makebox(0,0)[lb]{\smash{{\SetFigFont{7}{8.4}{\rmdefault}{\mddefault}{\updefault}$\pi_i$$\ldots\pi_{j-1}$        }}}}
\put(3001,-4036){\makebox(0,0)[lb]{\smash{{\SetFigFont{7}{8.4}{\rmdefault}{\mddefault}{\updefault}$\pi_1$$\ldots\pi_{i-1}$}}}}
\put(601,-4486){\makebox(0,0)[lb]{\smash{{\SetFigFont{7}{8.4}{\rmdefault}{\mddefault}{\updefault}Scenario 4}}}}
\put(526,-5686){\makebox(0,0)[lb]{\smash{{\SetFigFont{7}{8.4}{\rmdefault}{\mddefault}{\updefault}Scenario 5}}}}
\put(6751,164){\makebox(0,0)[lb]{\smash{{\SetFigFont{7}{8.4}{\rmdefault}{\mddefault}{\updefault}$\tau_1$}}}}
\put(6751,-1111){\makebox(0,0)[lb]{\smash{{\SetFigFont{7}{8.4}{\rmdefault}{\mddefault}{\updefault}$\tau_2$}}}}
\put(6751,-2461){\makebox(0,0)[lb]{\smash{{\SetFigFont{7}{8.4}{\rmdefault}{\mddefault}{\updefault}$\tau_3$}}}}
\put(6751,-3811){\makebox(0,0)[lb]{\smash{{\SetFigFont{7}{8.4}{\rmdefault}{\mddefault}{\updefault}$\tau_4$}}}}
\end{picture}%

%% file: combined.bbl
\begin{thebibliography}{10}

\bibitem{amm}
Dweighter, H.
\newblock American Mathematical Monthly 82(1975), 1010

\bibitem{BS03}
Bass, D.W., Sudborough, I.H.:
\newblock Pancake problems with restricted prefix reversals and some
  corresponding cayley networks.
\newblock Journal of Parallel and Distributed Computing archive \textbf{63}(3)
  (March 2003)  327--336

\bibitem{HP95}
Hannenhalli, S., Pevzner, P.:
\newblock Transforming cabbage into turnip.
\newblock In: Proc. of 27th Annual ACM Symposium on Theory of Computing
  (STOC'95). (1995)  178--189

\bibitem{HS93}
Heydari, M.H., Sudborough, I.H.:
\newblock On sorting by prefix reversals and the diameter of pancake networks.
\newblock Parallel Architectures and Their Efficient Use \textbf{678} (1993)
  218--227

\bibitem{HS97}
Heydari, M.H., Sudborough, I.H.:
\newblock On the diameter of the pancake network.
\newblock Journal of Algorithms \textbf{25}(1) (1997)  67--94

\bibitem{KS93}
Kececioglu, J., Sankoff, D.:
\newblock Exact and approximation algorithms for the inversion distance between
  two permutations.
\newblock In: Proc. of 4th Annual Symposium on Combinatorial Pattern Matching,
  (CPM'93). Volume 684 of Lecture Notes in Computer Science., Springer (1993)
  87--105 Extended version has appeared in {\it Algorithmica}, 13:180-210,
  1995.

\bibitem{BP93}
Bafna, V., Pevzner, P.:
\newblock Genome rearrangements and sorting by reversals.
\newblock In: Proc. of 34th Annual IEEE Symposium on Foundations of Computer
  Science (FOCS'93). (1993)  148--157 Also in {\it SIAM Journal on Computing},
  25:272-289, 1996.

\bibitem{BHK02}
Berman, P., Hannenhalli, S., Karpinski, M.:
\newblock 1.375-approximation algorithm for sorting by reversals.
\newblock In: Proc. of 10th European Symposium on Algorithms (ESA'02). Volume
  2461 of Lecture Notes in Computer Science., Springer (2002)  200--210

\bibitem{BI96}
Christie, D.A.:
\newblock Sorting permutations by block-interchanges.
\newblock Information Processing Letters \textbf{60}(4) (1996)  165--169

\bibitem{BP95}
Bafna, V., Pevzner, P.:
\newblock Sorting permutations by transpositions.
\newblock In: Proc. of the 6th Annual ACM-SIAM Symposium on Discrete Algorithms
  (SODA'95). (1995)  614--623 Also in {\it SIAM Journal on Discrete
  Mathematics}, 11(2):224-240, 1998.

\bibitem{EH05}
Elias, I., Hartman, T.:
\newblock A 1.375-approximation algorithm for sorting by transpositions.
\newblock In: Proc. of 5th International Workshop on Algorithms in
  Bioinformatics ({WABI}'05). Volume 3692 of Lecture Notes in Computer
  Science., Springer (October 2005)  204--214

\bibitem{H03}
Hartman, T.:
\newblock A simpler 1.5-approximation algorithm for sorting by transpositions.
\newblock In: Proc. of 14th Annual Symposium on Combinatorial Pattern Matching
  (CPM'03). Volume 2676 of Lecture Notes in Computer Science., Springer (2003)
  156--169

\bibitem{HS04}
Hartman, T., Sharan, R.:
\newblock A 1.5-approximation algorithm for sorting by transpositions and
  transreversals.
\newblock In: Proc. of 4th International Workshop on Algorithms in
  Bioinformatics (WABI'04). Volume 3240 of Lecture Notes in Computer Science.,
  Springer (2004)  50--61

\bibitem{FF06}
Lu, C.L., Huang, Y.L., Wang, T.C., Chiu, H.T.:
\newblock Analysis of circular genome rearrangement by fusions, fissions and
  block-interchanges.
\newblock BMC Bioinformatics (12 June 2006)

\bibitem{DM02}
Dias, Z., Meidanis, J.:
\newblock Sorting by prefix transpositions.
\newblock In: 9th International Symposium on String Processing and Information
  Retrieval (SPIRE 2002). Volume 2476 of Lecture Notes in Computer Science.,
  Springer (2002)  463--468

\bibitem{C97}
Caprara, A.:
\newblock Sorting by reversals is difficult.
\newblock In: Proc. of 1st ACM Conference on Research in Computational
  Molecular Biology (RECOMB'97). (1997)  75--83

\bibitem{HS}
Heydari, M.H., Sudborough, I.H.:
\newblock Sorting by prefix reversals is np-complete To be submitted (as
  mentioned in~\cite{W98}).

\bibitem{W98}
Walter, M., Dias, Z., Meidanis, J.:
\newblock Reversal and transposition distance of linear chromosomes.
\newblock In: South American Symposium on String Processing and Information
  Retrieval (SPIRE'98), IEEE Computer Society (1998)  96--102

\bibitem{GPS99}
Gu, Q., Peng, S., Sudborough, H.:
\newblock A 2-approximation algorithm for genome rearrangements by reversals
  and transpositions.
\newblock Theoritical Computer Science \textbf{210}(2) (1999)  327--339

\bibitem{LX99}
Lin, G., Xue, G.:
\newblock Signed genome rearrangements by reversals and transpositions: Models
  and approximations.
\newblock In: Proc. of 5th Annual International Conference on Computing and
  Combinatorics (COCOON'99). Volume 1627 of Lecture Notes in Computer Science.,
  Springer (1999)  71--80

\bibitem{E02}
Eriksen, N.:
\newblock (1+$\epsilon$)-approximation of sorting by reversals and
  transpositions.
\newblock Theoretical Computer Science \textbf{289}(1) (2002)  517--529

\bibitem{Atif08}
Rahman, A., Shatabda, S., Hasan, M.:
\newblock An appoximation algorithm for sorting by reversals and
  transpositions.
\newblock Journal of Discrete Algorithms \textbf{6}(3) (2008)  449--457

\bibitem{LZ08}
Lou, X., Zhu, D.:
\newblock A 2.25-approximation algorithm for cut-and-paste sorting of unsigned
  circular permutations.
\newblock In: COCOON. (2008)  331--341

\end{thebibliography}
